\documentclass[conference]{IEEEtran}
\IEEEoverridecommandlockouts
\usepackage{cite}
\usepackage{amsmath,amssymb,amsfonts}
\usepackage{algorithmic}
\usepackage{graphicx}
\usepackage{textcomp}
\usepackage{xcolor}
\UseRawInputEncoding
\def\BibTeX{{\rm B\kern-.05em{\sc i\kern-.025em b}\kern-.08em
    T\kern-.1667em\lower.7ex\hbox{E}\kern-.125emX}}
\begin{document}

\title{Reconfigurable Intelligent Surface Assisted Railway Communications: A survey 
\
}

\author{\IEEEauthorblockN{Aline~Habib\IEEEauthorrefmark{1}, Ammar~El Falou\IEEEauthorrefmark{2}, Charlotte Langlais\IEEEauthorrefmark{1}, Marion Berbineau\IEEEauthorrefmark{4}}\\
\IEEEauthorblockA{\IEEEauthorrefmark{1} Mathematical and electrical engineering department, CNRS UMR 6285 Lab-STICC, IMT Atlantique, Brest, France}

 \IEEEauthorblockA{ \IEEEauthorrefmark{2} CEMSE Division, King Abdullah University of Science and Technology (KAUST), Saudi Arabia}
  \IEEEauthorblockA{ \IEEEauthorrefmark{4} COSYS-LEOST, {Université Gustave Eiffel, Villeneuve d'Ascq, France}\\
Email: \{aline.habib, charlotte.langlais\}@imt-atlantique.fr, ammar.falou@kaust.edu.sa, marion.berbineau@univ-eiffel.fr}

}
\maketitle

\begin{abstract}
The number of train passengers and the demand for high data rates to handle new technologies such as video streaming and IoT technologies are continuously increasing. Therefore the exploration of millimeter waves (mmWave) band is a key technology to meet this demand. However, the high penetration loss makes mmWave very sensitive to blocking, limiting its coverage area. One promising, efficient, and low-cost solution is the reconfigurable intelligent surface (RIS). This paper reviews the state of the art of RIS for railway communications in the mmWave context. First, we present the different types of RIS and review some optimization algorithms used in the literature to find the RIS phase shift. Then, we review recent works on RIS in the railway domain and provide future directions.
\end{abstract}


\section{Introduction}\label{Intro}

The need to double the capacity of the existing rail networks and, at the same time to increase the overall quality of service is leading to a drastic increase in the need for high data rates and robust and low latency data exchange between the different actors in the rail system. This multiplication of transmission needs ultimately leads to problems of spectrum scarcity. In this context, using mmWave bands opens up new opportunities. However, mmWaves  suffer from very high attenuation and high sensitivity to various masking effects. In this context, Reconfigurable Intelligent Surfaces offers promising application use cases.

Reconfigurable Intelligent Surface, known in the literature by several nomenclatures as  Software-Controlled Metasurface \cite{liaskos2018new}, Intelligent Reflecting Surface (IRS) \cite{9326394}, Large Intelligent Surface (LIS) \cite{8644519}, and Reconfigurable Smart Surface (RSS) \cite{alfattani2021link},  is an electromagnetic-based reconfigurable structure
 that turns the random nature of the propagation channel into a controllable and programmable radio environment. RIS is a thin planar meta-surface made of several low-cost reflective elements \cite{8910627}. Each RIS element adjusts the phase and amplitude of the incident wave to reflect it into a beam toward the target direction. This improves the signal quality and extends the coverage area especially when the direct link is blocked. The paper's main objective is to provide the reader with the basic elements to understand RIS and its interest in a railway communication environment. To do so, we review the literature in the domain and propose some future research directions.
 
  
The rest of the paper is organized as follows. Section~\ref{sec:RIS} provides a literature overview related to RIS, such as the different RIS structures and types, and their opportunity in the context of mmWave communications. We stress the need for realistic channel models in order to properly evaluate the performance of RIS-assisted systems. Section~\ref{sec:RIS_Railway} focuses on very recent  works investigating RIS-assisted systems for railway communications. Finally, in Section~\ref{sec:future}, some future directions are drawn, and Section~\ref{sec:conclusion} concludes the paper.

 \section{Reconfigurable Intelligent Surface}
 \label{sec:RIS}
 \subsection{RIS General Overview}
 \begin{figure}
\vspace{-0.4cm}
	\centering
\includegraphics[width=1.3\columnwidth, trim=180 160 40 70, clip]{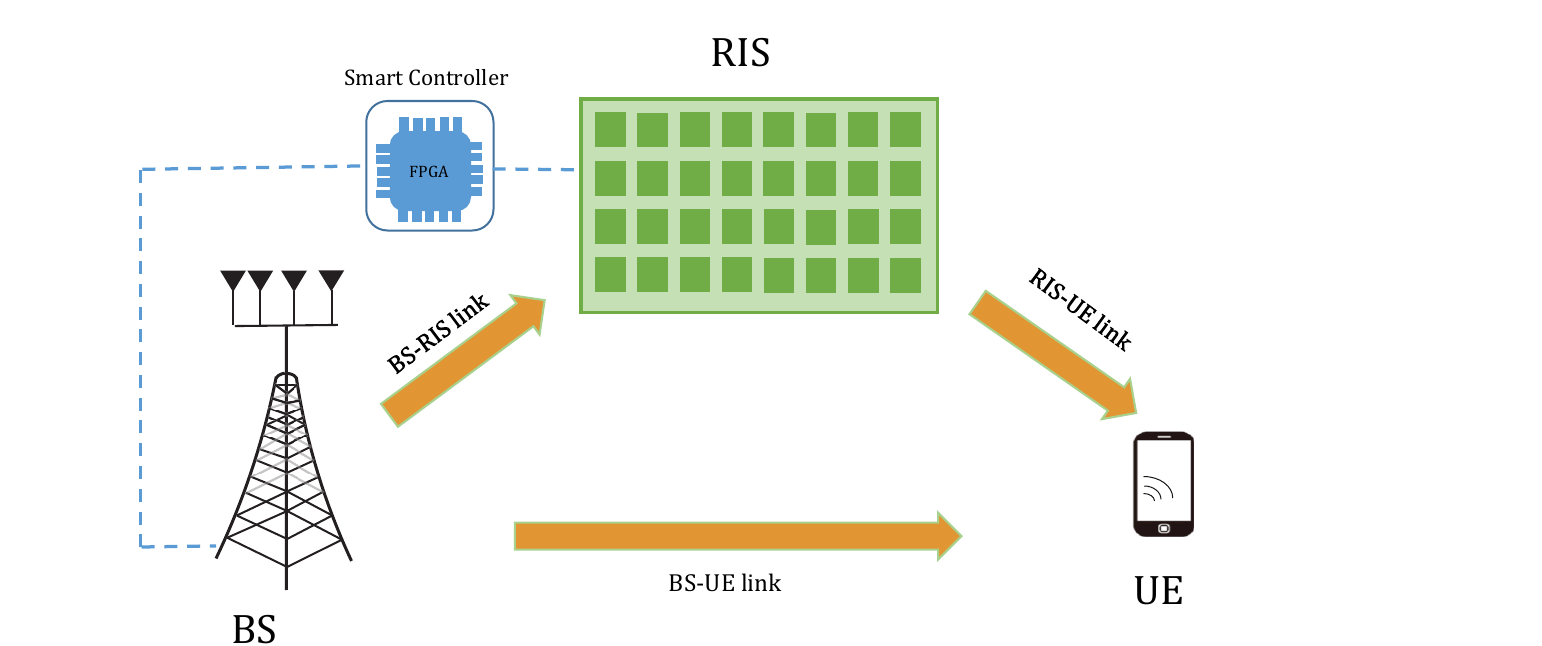}
	\caption{ RIS-assisted Single User MIMO system. 
 }
        \vspace{-0.2cm}
	\label{fig:ris}
	\vspace{-0.2cm}
\end{figure}
The main objective of a RIS is to provide a programmable radio environment between a transmitter (Tx), typically a base station (BS) in the downlink case, and a receiver (Rx), typically a remote user equipment (UE),  by changing the phase shifts and amplitude of the RIS incident wave as follows \cite{liu2021reconfigurable}\begin{equation}\label{coef}
  z_n=\beta_ne^{j\theta_n}, 
  \end{equation}
where $z_n$ is the reflection coefficient of the $n^{th}$ element,  $\beta_n$ and $\theta_n$ are the adjustments in amplitude and phase due to the $n^{th}$ element.
As the RIS should not encompass too many RF and signal processing resources to maintain a low level of energy consumption and complexity, the BS computes the needed tunable parameters and transfers commands to each RIS element thanks to a smart controller \cite{9241881}  as seen in Fig.\ref{fig:ris}.

To adjust phase shifts and amplitude of the incident wave, RIS consists of adjustable components, such as diodes and liquid crystals. The diodes adjust the signal by changing the bias voltage, while the liquid crystals adjust the electromagnetic signal by changing material parameters such as conductivity and permeability \cite{9828501}. Indeed, the PIN diode-based RIS consists of three layers: 1) The outer layer with printed metal patches on a dielectric substrate. This layer directly processes the incident signals. 2) The intermediate layer
composed of a copper panel to avoid signal energy loss. 
3) The inner layer is a control board activated by a programmable digital electronic circuit  (FPGA), allowing the real-time adjustment of the  RIS elements' reflection coefficients \cite{9828501}.

Two reflexion paradigms govern propagation in the context of RIS-assisted communication systems, namely, the specular reflection paradigm and the scattering reflection paradigm,  \cite{alfattani2021link}. The differences are mainly related to the relation between the size of RIS $A_t$ and the distance D between BS-RIS or RIS-UE, as follows:
\begin{itemize}
    \item The specular reflection paradigm: the transmission occurs in the near-field, i.e., $D<d_{lim} = \frac{2A_{t}}{\lambda}$\footnote{$d_{lim}$ denotes the Rayleigh distance and is defined by $d_{lim}=\frac{2A_{t}}{\lambda}  
    $ with $A_{t}$ the RIS area and $\lambda$ the wavelength \cite{alfattani2021link}.}. The path loss, in this case, depends on the summation of the distances between BS-RIS and RIS-UE. 
    \item The scattering reflection paradigm: the transmission occurs in the far-field, i.e., $D>d_{lim}$. In this case, the path loss depends on the product of the BS-RIS and RIS-UE separation distances.
\end{itemize}

\begin{figure}
	\centering
	\includegraphics[width=1.25\columnwidth, trim=90 268 15 275, clip]{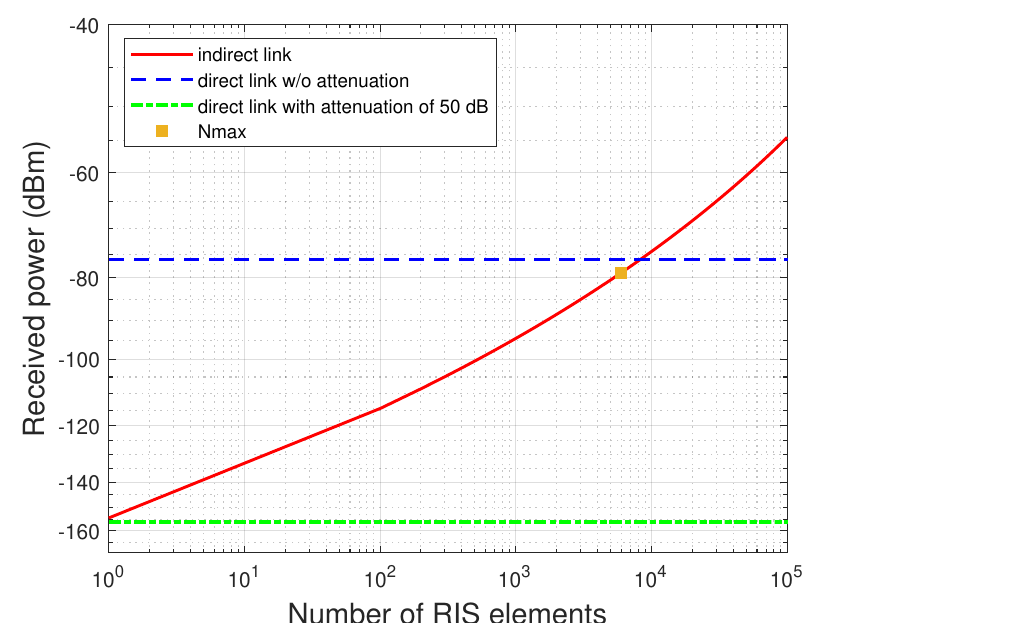}
	\caption{Received power versus the number of RIS elements for the direct link (distance from BS to UE equal to $158.7$~m) with an obstacle of  50~dB attenuation  and without obstacles and the indirect link via RIS ($d_1= 90$~m; $d_2 = 30$~m). }  
   \label{direct_indirect_Pr }
   \vspace{-0.35cm}
\end{figure}

In the case of a passive RIS ($\beta_n \leq 1$),  the RIS elements reflect the signal without amplification. Thus, in the context of scattering reflection  communications (far-field), and by assuming the optimal phase shifts, the received power at the UE for the indirect link via passive  RIS is expressed as \cite{alfattani2021link}
\begin{equation}\label{eq:pN}
P_{r}^{\text{UE}}=P_tG_tG_r\left(\frac{\lambda}{4\pi}\right) ^4\frac{(d_0)^{2\mu-4}}{{(d_1d_2)^\mu}}N^2 
\end{equation}
where $P_t$ is the transmitted power at the BS, $G_t$ and $G_r$ are the transmit and receive antenna gains at the BS and the UE, respectively, $d_0$ the reference distance in the free space, $d_1$ and  $d_2$ are  BS-RIS distance and  RIS-UE distance, $\mu$ is the path loss exponent depending on the environment type (e.g., $\mu \ge 3$ for urban environments), and $N$ is the number of  RIS elements.  Thus, the passive RIS gives a gain proportional to $N^2$. However, the passive RIS has limitations due to the double-path loss effect. Indeed, the signal traverses two cascaded channels,  the Tx-RIS link and the RIS-Rx  link \cite{9998527}. Thus, the received power via the indirect link could be greater
than the power of the direct link, if $N$ is large, or/and if
the direct link is weak or blocked. To illustrate this concept, we plot in  Fig.  \ref{direct_indirect_Pr } the power received at the UE via an
attenuated direct link, a direct link without attenuation, and the indirect
link via the RIS, versus the number of RIS elements.  The mmWave channel links are generated using an extended version of the New York University simulator NYUSIM \cite{habib2023extended}.  Note that to verify the RIS scattering reflection paradigm, the distances $d_1$, and $d_2$ must be in the far-field region. As  RIS size increases, the distance where the RIS is in the near-field also increases. Thus, for the distances $d_1$, and $d_2$ to be in the far-field and equation (\ref{eq:pN}) to be valid,   $N$ must not exceed a certain $N_{\max}$, computed from $d_{lim}$, the Rayleigh distance, and represented by a square in Fig. \ref{direct_indirect_Pr } \cite{habib2023extended}.  The behavior of the RIS in the near-field  is an interesting research topic.

\subsection{RIS types}
 To overcome this limitation and obtain an efficient RIS when the direct link exists or the number of RIS elements is low, the authors of \cite{9998527} propose an active RIS that can amplify the reflected signals through amplifiers embedded in the RIS elements.  The simulation results in a direct link scenario without attenuation for 256 RIS elements   reveal  a negligible sum-rate gain 
of $3~\%$ using the passive RIS, while their proposed active RIS offers a significant sum-rate gain   of $67~\%$ compared to the case without RIS.

Nevertheless, a RIS with a large number of active elements consumes more energy. Thus, the authors in \cite{9833956}  propose a novel type of RIS composed of active and passive reflective elements, called hybrid RIS, to deal with the limited power budget of the RIS.

RIS based on continuous phase shifts is considered an ideal system that is difficult to implement in practice. Therefore, RIS based on finite discrete phase shifts is the alternative solution to cope with this hardware constraint. To this end, the authors in \cite{9223720} compare the performance of RIS systems with continuous and discrete phase shifts and they find that 3 levels of quantization are sufficient to obtain full diversity.

%


\subsection{RIS optimization}
The efficient functioning of the RIS is strongly affected by the adapted phase shifts $\theta_n$. For instance, in Single Input Single Output (SISO)  systems, the optimal phase shift of a RIS is  easily determined analytically as follows \cite{alfattani2021link}
\begin{equation}\label{eq:angle_opt}
\theta_n=\theta_{tn}+\theta_{nr}. 
\end{equation} where $\theta_{tn}$ and 
$\theta_{nr}$ are the phase of the LoS path in the BS-RIS
and RIS-UE channels, respectively.

However, it is hard to find the optimal phase shifts analytically in the case of Multiple Input Multiple Output (MIMO) systems. To this end, an optimization algorithm is needed. \cite{8982186} studied multi-user Multi Input Single Output (MISO) downlink communications assisted by RIS, where the objective is to maximize the weighted sum rate to find  the optimized passive beamforming $\theta_n$ and the optimized precoding at the BS.
To solve this non-convex problem, they used the  Lagrangian Dual Transform which transforms the sum-of-logarithms-of-ratio to an alternative form. \\
The authors in \cite{9606297} discussed an indoor MISO multi-user system with a channel model based on the Rician K-factor. The RIS phase shift were configured as follows 
\begin{equation}
   \theta_n^*=\arg(\textbf{H}_d^H) - \arg(\textbf{H}_l^H)-\arg(\textbf{H}),
\end{equation}
where $\textbf{H}_d$ is the direct channel between the Tx and the Rx, $\textbf{H}$ is the channel  between Tx and RIS, and $\textbf{H}_l$ is the channel between the RIS and the $l$th user. 

In \cite{9160972}, the authors adopted a low-complex algorithm called  the cosine similarity algorithm. The latter  aims to find the sub-optimal phase shifts of the RIS that maximize the channel gain.  Moreover, to minimize the transmitted power given the bit error rate for a RIS-assisted single-user multipath uplink system,
the authors of \cite{9495932} propose an iterative algorithm to jointly optimize precoding and passive beamforming. In addition, a deep learning algorithm is  applied in \cite{khan2019deep} to maximize the received signal-to-noise ratio and  find the optimal phase shifts of RIS. 


\subsection{RIS versus Relay}
Both RIS and relay aim to improve signal quality and coverage. However, there are two main differences.
\begin{itemize}
    \item In the case of RIS, a power supply is only needed to configure the RIS components based on low-cost materials (diodes, switches...). Once the configuration is done, the RIS becomes passive, and no power supply is needed \cite{bjornson2020reconfigurable}. However, relays are generally considered active devices connected to active electronics such as analog-to-digital converters, digital-to-analog converters, amplifiers, etc., which require a power supply for operation. As a result, relays are more complex to implement and consume more energy than RIS \cite{9119122}.
    \item A RIS operates in a full duplex mode while  relays generally work in a half-duplex mode. Relays can still operate  in full duplex mode, but this increases their cost, since appropriate antennas and analog and/or digital signal processing, to eliminate loop-back self-interference, are required \cite{9119122}.

\end{itemize}



\subsection{RIS is an opportunity for mmWave communications}
The mmWave band, ranging from 30 to 300~GHz, offers enormous free bandwidth and high data rate possibilities  \cite{9350499}, unlike the overloaded low-frequency spectrum. However, it is very vulnerable to oxygen absorption and rain attenuation, and also suffers from penetration loss that makes mmWave signals easily blocked. Therefore, the coverage of mmWave communications is limited\cite{9568459}. On the other hand,  when the direct link is blocked or largely attenuated, a RIS is a competitive solution to extend  coverage area and connectivity \cite{9201413}. The location of the RIS should be optimized to obtain two efficient connections: the BS-RIS link and the RIS-UE link. 

The authors in \cite{9824117} discuss the size limitation of the RIS in low frequencies below 6~GHz, which makes their deployment in this band inefficient. A study of the specific propagation characteristics of the terahertz  band is needed to use RIS in these frequencies, and the most important implementation of RIS today is in the mmWave band. 
 
In the literature, the most used channels in RIS-assisted systems are the theoretical channels  such as Rice for Line-of-Sight (LOS) environments, and Rayleigh for non-LOS (NLOS) \cite{8982186}, \cite{9738798}. To fill the gap towards realistic channel modeling and simulator, the authors in \cite{9397266}  propose a novel geometrical  channel simulator, called SimRIS. This simulator is based on statistical modeling and can be used in indoor and outdoor environments at 28 and 73 GHz frequencies. Moreover, in \cite{dorokhin2022reconfigurable}
the authors extend QuaDRiGa, a simulator used to model MIMO radio channels at sub-6GHz and mmWave frequencies,  to handle RIS. This simulator  is convenient for RIS-assisted MIMO systems with  a  mobile Rx or mobile RIS. In addition, \cite{habib2023extended} discusses the extension of NYUSIM, a mmWave channel simulator based on extensive measurements and well-used to assess MIMO systems \cite{el2018performance,khaled2019performance,khaled2021multi}, to generate realistic channels for RIS-assisted systems.


\section{RIS-assisted railway communications}
\label{sec:RIS_Railway}

\subsection{Railway environments characteristics}  

Railway environments are known to be very complex and harsh from a radio point of view. Various obstacles such as pylons supporting the catenary and rapid transitions between different scenarios (cutting/tunnel, cutting/viaduct) can create severe radio impairments. Railway tunnel size and shape are very specific, depending on the category of the train. Radio propagation inside tunnels is often modeled using Ray tracing tools \cite{Qiu2021}, \cite{He2018}. It is also important to mention that MIMO system performance in tunnels is subject to possible impairments depending on spatial correlation in the tunnel and also Key holes phenomenon \cite{Berbineau2021}. Due to high speed, the train can rapidly go through diverse scenarios. In addition, Doppler effects and possible interference due to the proximity of high voltage (catenary) in the vicinity of the antennas render the railway environments very specific compared to the indoor, urban, or suburban environments generally considered today for the use of mmWave communication systems. A detailed description of railway-specific environments can be found in \cite{Berbineau2021}.

Considering the capability of RIS to solve the blockage problems in mmWave wireless communications, the use of RIS for railway communications has recently been considered as a promising candidate.

\subsection{RIS-assisted railway communications}
\subsubsection{RIS for high-speed trains}
\cite{Xu2022-2} discusses the need for RIS in High-Speed Railway (HSR) environment for mmWave communications to improve the signal quality, which suffers from frequent blockages due to high-speed trains. The authors apply Deep reinforcement Learning (DRL) based approach to jointly optimize the RIS phase shifts and the BS beamforming for spectral efficiency maximization. The results show a significant improvement in spectral efficiency performance using DRL compared to the traditional approach. 

In \cite{9690475}, the authors describe how to use RIS on high-speed trains to improve communication performance by providing beamforming, interference mitigation, and reducing signal attenuation. They present a detailed discussion of the challenges associated with the RIS deployment on these trains, such as the need for tracking of the train, low latency, and high-speed RIS control, and the impact of train vibration on the RIS performance. They also propose the DRL approach to solve the sum rate maximization problem.

\cite{9814619} deals with interference suppression in an HSR network,  composed of a BS, a mobile relay (MR) located on the train, a RIS located near the MR, and an interference
source. The authors maximize the channel capacity 
using a DRL solution and they consider outdated channel state information (CSI) to take into account the motion of the train. The authors found that deploying a RIS in close proximity to the embedded  MR improves interference suppression and that their algorithm is more effective in suppressing interference than other optimization algorithms
based on mathematical formulations.

\cite{9933649}  proposes a new interrupt flow scheduling approach for RIS-assisted downlink mmWave HSR communications where multiple mobile relays exist. Given the existence of eavesdroppers, the BS schedules a  number of flows for each MR when the MR flow quality of service (QoS) exceeds the QoS requirement. The authors seek to maximize the scheduled flow number, find the optimal beamforming, the optimal RIS phase shifts, and the scheduling or not of the RIS discrete phase shift, and they find that RIS can intend communication security by reducing eavesdropping capacity and extending coverage area in the HSR environments.


\subsubsection{RIS in railway tunnels}
In \cite{Chen2022}, the authors have considered a simple two dimensions empty tunnel. Using the image theory approach and a vertical blocking element between a Tx and an Rx inside the tunnel, they have shown that the use of RIS located on the ceiling of the tunnel can reduce the Blocking Probability (BP) of the signal between Tx 
and Rx. 
An increase in the number of RIS 
and optimization of the Tx position conduct to an additional decrease in BP. The increase in
distance between RIS and Tx can extend the effective range of RIS for a given BP. This study could be extended by considering a train inside a 3D tunnel.

\subsubsection {RIS for passengers inside trains}
  Recently RIS technology has been studied to extend the coverage area in the mmWave band inside an airplane cabin \cite{li2023mmwave}. The authors aim to  minimize the number of RIS deployed in this system while ensuring the user data rate remains above a threshold. Besides, they compare the performance of this system for two RIS positions in the cabin corridor near the seat and above the center seat. This study could be easily transposed to the case of the inside of a high-speed train or inside a metro to guarantee a given throughput for the passengers.




\section{Future directions}
\label{sec:future}
As discussed in the previous sections, RIS offers a promising low-cost solution to solve the blocking problems in railway networks since it improves the efficiency and reliability of high-speed trains, solves the interference problem, and extends the coverage area through controlled signal reflection. In the case of high-speed trains, the channel estimation for RIS-assisted communications is a crucial challenge due to the unexpected rapid change of environments. Future research directions could explore the case of RIS-assisted wireless communications in tunnels, especially when the vertical cross-section of the train is large compared to the tunnel cross-section, which increases the probability of signal blockage. In addition, the case where the train moves in the tunnel from the inside to the outside is particularly difficult due to the development of urban transport and in particular driverless metro systems which require high data rate transmissions. The optimization of RIS-assisted communications in this case will require the development of realistic channel models. It would also be interesting to study the optimal location of the RIS, the number of RIS elements, or the number of RIS itself, needed in these systems to maximize the coverage inside the tunnel and also maximize the ever-increasing passenger throughput demand onboard the trains.

\section{Conclusion}
\label{sec:conclusion}
This paper presents a survey on RIS-assisted communications for railway applications, particularly in the mmWave band. First, we have defined the RIS concept, explaining its structure, and different types of RISs. A review of the various optimization algorithms used in the literature for RIS-assisted systems is proposed, and we highlight the ability of RIS to solve the blocking problem of mmWave. In the last section, the paper outlines
the characteristics of the railway environments and details some recent works concerning the use of RIS in high-speed trains. This topic is a very active field of research and we have proposed some future directions  for RIS-assisted railway communications.

\section*{Acknowledgment}
This work was funded by the council of the Region Bretagne, under the grant MILLIRIS.
\bibliographystyle{IEEEtran}
\bibliography{Bibliography}

\vspace{12pt}

\end{document}